\documentclass[12pt]{iopart}
\usepackage{graphicx}% Include figure files

\usepackage{iopams}
\usepackage{graphicx}
\usepackage{dcolumn}
\usepackage{bm}
\usepackage{float}
\usepackage{cite}
\expandafter\let\csname equation*\endcsname\relax
\expandafter\let\csname endequation*\endcsname\relax
\usepackage{amsmath}
\usepackage{amssymb}
\usepackage[pdfstartview=FitH, bookmarksnumbered=true, bookmarksopen=true, colorlinks=true, pdfborder=001, citecolor=blue, linkcolor=blue, linktocpage=true] {hyperref}

\begin{document}

\title{Topological pumping induced by spatiotemporal modulation of interaction}

\author{Boning Huang $^{1,2}$, Yongguan Ke $^{1,2,4,*}$, Wenjie Liu $^{5}$, Chaohong Lee $^{2,3,5,\dag}$}

\address{$^{1}$Laboratory of Quantum Engineering and Quantum Metrology, School of Physics and Astronomy, Sun Yat-Sen University (Zhuhai Campus), Zhuhai 519082, China}

\address{$^{2}$Institute of Quantum Precision Measurement, State Key Laboratory of Radio Frequency Heterogeneous Integration, Shenzhen University, Shenzhen 518060, China}

\address{$^{3}$College of Physics and Optoelectronic Engineering, Shenzhen University, Shenzhen 518060, China}

\address{$^{4}$International Quantum Academy, Shenzhen 518048, China}

\address{$^{5}$Quantum Science Center of Guangdong-Hongkong-Macao Greater Bay Area (Guangdong), Shenzhen 518045, China}

\ead{$^*$keyg3@mail.sysu.edu.cn}
\ead{$^\dag$chleecn@szu.edu.cn}

\begin{abstract}
Particle-particle interaction provides a new degree of freedom to induce novel topological phenomena. 
Here, we propose to use spatiotemporal modulation of interaction to realize topological pumping without single-particle counterpart.
%
%In the absence of interaction, the particles independently hop in the lattice, which is completely trivial.
%
%However, when two particles are at the same site,
%
%strong interaction makes the two particles as a bound state, which also feels a spatiotemporal potental.
%
Because the modulation breaks time-reversal symmetry,
the multiparticle energy bands of bound states have none-zero Chern number, and support topological bound edge states.
In a Thouless pump, a bound state that uniformly occupies a topological energy band can be shifted by integer unit cells per cycle, consistent with the corresponding Chern number.
We can also realize topological pumping of bound edge state from one end to another.
The entanglement entropy between particles rapidly increases at transition points, which is related to the spatial spread of a bounded pair. 
In addition, we propose to realize hybridized pumping with fractional displacement per cycle by adding an extra tilt potential to separate topological pumping of the bound state and Bloch oscillations of single particle.
Our work could trigger further studies of correlated topological phenomena that do not have a single-particle counterpart. 
\end{abstract}

\vspace{2pc}

\maketitle

\section{Introduction}
Topological pumping, topologically protected transport of particles assisted by the topological structure of spatiotemporal modulation, has attracted intense interest and attention in recent years~\cite{citro2023thouless}.
Topological pumping can be mainly classified as quantized Thouless pumping~\cite{PhysRevB.27.6083}, nonquantized geometrical pumping~\cite{PhysRevLett.116.200402}, and topological pumping of edge states~\cite{PhysRevLett.109.106402,cheng2022asymmetric}.   
In Thouless pumping, the displacement of an initial state uniformly filling a topological band is related to the nontrivial topological invariant known as the Chern number~\cite{PhysRevB.47.1651,RevModPhys.82.1959}.
However, in geometrical pumping only a single-momentum state of the topological band is involved, and the displacement is determined by the nonquantized Berry phase~\cite{PhysRevLett.116.200402}.
According to bulk-edge correspondence, topological edge states exist in the band gap if the topological invariant of the below bands is nontrivial, providing a topological pumping channel from one edge to the other edge.
Due to the rapid development of quantum technologies~\cite{PhysRevLett.111.026802,PhysRevA.90.063638,PhysRevA.92.013609,https://doi.org/10.1002/lpor.201670069}, topological pumping has been experimentally realized in a variety of quantum simulators~\cite{PhysRevLett.109.106402,PhysRevB.91.064201,lohse2016thouless,nakajima2016topological,PhysRevLett.116.200402,cerjan2020thouless,grinberg2020robust,PhysRevLett.126.095501,jurgensen2021quantized,cheng2022asymmetric,sun2022non,You2022,jurgensen2023quantized}.
%Boosted by experimental progresses, theories of Thouless pumping are generalized to topological systems involved with disorder, non-Abelian, interaction,  higher-order topological phases. 
%
 
Interactions between particles cause novel correlation and entanglement, which may introduce rich correlated topological phenomena~\cite{Gorlach2017,PhysRevA.95.063630,PhysRevB.96.195134,PhysRevB.98.245148,vanVoorden_2019,olekhno2020topological,PhysRevA.102.013510,PhysRevResearch.2.042024,PhysRevResearch.2.033267,PhysRevResearch.2.033190,PhysRevResearch.2.033143,azcona2021doublons,Zheng_2023,PhysRevB.107.125161,PhysRevResearch.5.013020}. 
Understanding the interaction effect in topological pumping could open a window to the largely unexplored field.
Although interactions break translational symmetry of individual particles, interacting systems may still preserve cotranslation sysmetry, that is, the energy is invariant if all particles as a whole are shifted by multiple unit cells~\cite{PhysRevA.95.063630,PhysRevB.96.195134,Qin_2018}.
With the help of cotranslation symmetry, single-particle Thouless pumping has been generalized to the multiparticle case~\cite{PhysRevA.95.063630,PhysRevA.101.023620,ke2023topological}.
Recently, Thouless pumping of bound state and topologically resonant tunneling have been experimentally realized in cold atomic systems~\cite{walter2023quantization} and superconducting qubits~\cite{tao2023interactioninduced}.
In particular, topologically resonant tunneling describes particles transported one-by-one by unit cells per cycle, which has no counterpart of single-particle topological pumping.
Up to now, interaction-induced topological pumping relies on spatiotemporal modulation of two more parameters, such as hopping and onsite potential, or hopping and interaction. 
It is still unclear whether multiparticle topological pumping can be induced by only modulating the interaction.     
Another fascinating topological phenomenon without single-particle counterpart is fractional topological pumping~\cite{Zeng2015,Zeng2016,Taddia2017}, in which long-range interaction is generally needed to realize fractional displacement per cycle.
It is highly appealing to search for a novel and simple way to realize fractional topological pumping.

In this paper, we predict multiparticle topological pumping of the bound bulk state and bound edge state in a Bose-Hubbard model with modulated interaction. 
With an initial state given by a maximally localized multiparticle Wannier state, the mean position shift in a pumping cycle will be determined by the Chern number of the filled band. 
Under open boundary conditions, there are edge modes and transport from one side to another can be performed. 
The evolution of the entanglement entropy between particles is also explored.
Without uniform occupation, the multiparticle Bloch states occupying only a single momentum correspond to a geomerical pumping with nonquantized charge pump. 
In addition, we propose a hybridized pumping with fractional mean position shift per cycle by combining the multiparticle Thouless pumping and Bloch oscillations of a single particle.
Our work provides a new manner to realize multiparticle Thouless pumping and hybridized fractional pumping, and brings a new sight to the interacting topological systems.

\section{Bose-Hubbard model with modulated interction}
\begin{figure}
	\centering
	\includegraphics[width=0.6\linewidth]{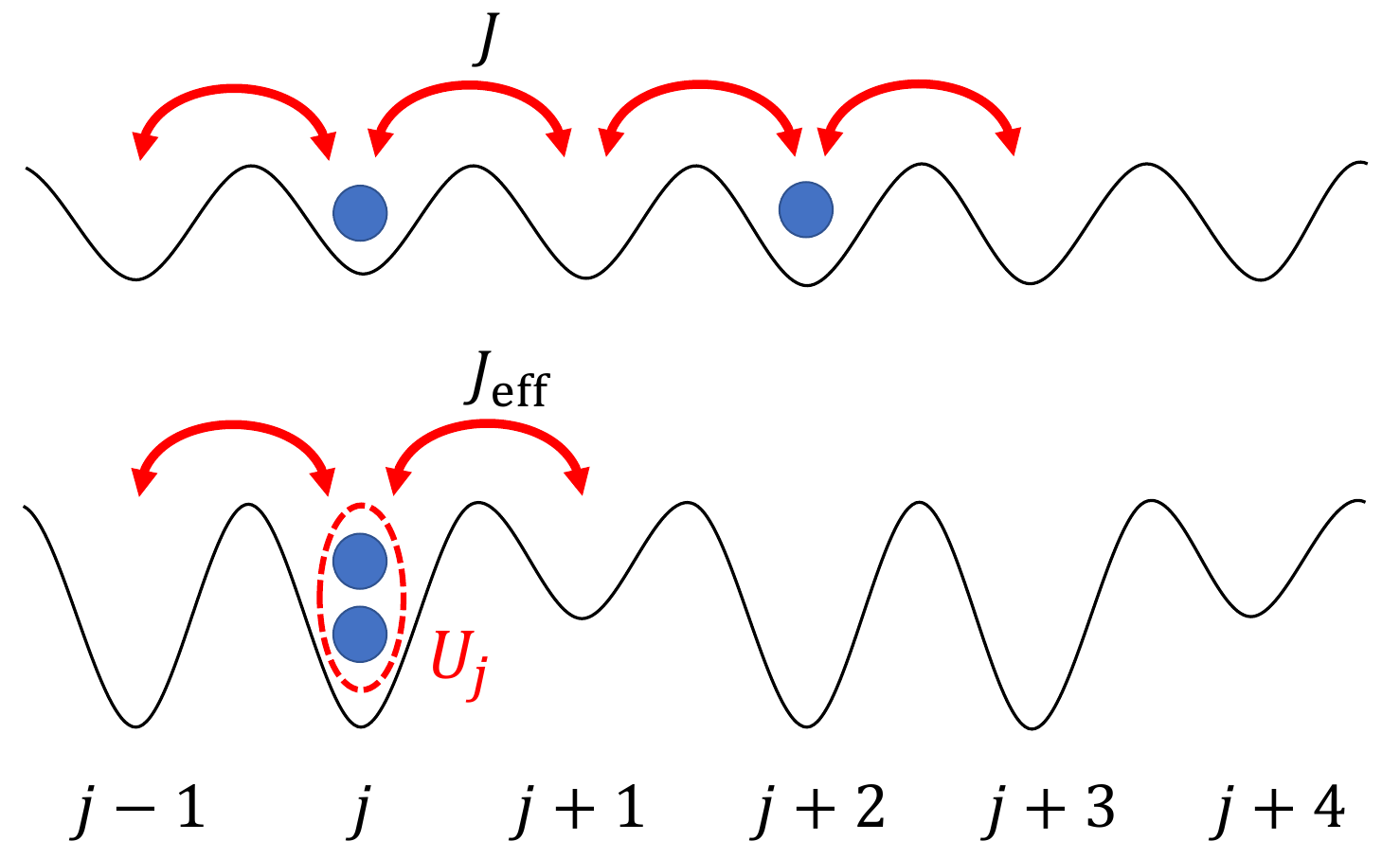}
	\caption{Schematics for two bosons in a lattice with spatiotemporal modulation of interaction.
    If the two bosons are separated, the interaction plays no role. If two bosons are at the same site, they can feel spatiotemporal modulation of onsite potential. }
	\label{fig:1}
\end{figure}
We consider $N$ bosons in a one dimensional lattice with $M$ sites, where the onsite interactions between bosons are periodically modulated, as depicted in Fig.~\ref{fig:1}.
The modulation of onsite interaction works on bosons at the same site but plays no role for a single boson or bosons at different sites. 
The system is governed by the following Hamiltonian,
\begin{equation}	
		\hat{H}_M=J\sum_{j}^{}(\hat{a}_{j+1}^{\dag}\hat{a}_j+h.c.)+\frac{1}{2}\sum_{j}^{}(U_0+\delta \cos(2\pi\beta j+\phi))\hat{n}_j(\hat{n}_j-1).	
\end{equation}
Here, $\hat{a}_{j}^{\dag}$ ($\hat{a}_j$) are the bosonic creation (annihilation) operators at the $j$th site. 
$\hat{n}_j=\hat{a}_{j}^{\dag}\hat{a}_j$ is the particle number operator at the $j$th site. 
$J$ is the nearest neighboring hopping strength, and $U_j=U_0+\delta \cos(2\pi\beta j+\phi)$ is the periodically modulated onsite interaction strength between the particles, where $\beta=p/q$ is the spatial modulation frequency and $\phi$ can be tuned temporally as $\phi=\omega t$ with modulation frequency $\omega$ and period $T=2\pi/\omega$.
$\delta$ is the modulation strength of the onsite interaction. 
The spatiotemporal modulation of interaction can be readily realized in some quantum simulators.
In optical lattices of cold atoms,
it is possible to tune the onsite interaction independently and dynamically through the Feshbach resonance ~\cite{walter2023quantization,PhysRevLett.116.205301}.
In arrays of superconducting circuits, modulation of the onsite interaction can be achieved by controlling the anharmonicity of individual qubits~\cite{tao2023interactioninduced}.
%
%Considering periodic boundary condition, When the index $j=M$, $j+1$ can be treated as $1$.

Since the modulation of interaction only takes effect on bound states, we focus on the subspace of two-particle bound state and derive an effective Hamiltonian of bounded pair through perturbation theory~\cite{bravyi2011schrieffer}. 
The hopping term
\begin{equation}
\hat{H}'=J\sum_{j=1}^M(\hat{a}^{\dag}_{j+1}\hat{a}_j+h.c.)
\end{equation}
can be treated as a perturbation to the interaction term
\begin{equation}
	\hat{H}_0=\frac{1}{2}\sum_{j=1}^M(U_0+\delta \cos(2\pi\beta j+\phi))\hat{n}_j(\hat{n}_j-1).
\end{equation} 
For convenience, considering two bosons, there are eigenstates $|2\rangle_j$ of $\hat{H}_0$ with eigenvalues $E_j=U_0+\delta \cos(2\pi\beta i+\phi)$ forming a subspace $\mathcal U$, and eigenstates $|1\rangle_j|1\rangle_k(j\neq k)$ with eigenvalues $E_{j,k}=0$ forming a complement subspace $\mathcal V$. Here, $|{2}\rangle_j$ are short for Fock states $|0,...,n_j=2,..., 0\rangle$ and $|1\rangle_j|1\rangle_k$ are short for $|0,...,n_j=1,...,n_k=1,..., 0\rangle$. 
The projection operators on $\mathcal U$ and $\mathcal V$ are respectively defined as
\begin{equation} \label{Projection}
	\begin{aligned}
		\hat{P}&=\sum_j|2\rangle_j\langle 2|_j,\\
		\hat{S}&=\frac{1}{2}\sum_{j\neq k}(\frac{1}{E_j-E_{j,k}}+\frac{1}{E_k-E_{j,k}})|1\rangle_j|1\rangle_k\langle 1|_k\langle 1|_j.
	\end{aligned}
\end{equation}
Applying the perturbation theory up to the second order~\cite{MTakahashi_1977,PhysRevA.90.062301}, the effective Hamiltonian of subspace $\mathcal U$ is given by
\begin{equation}	
\hat{H}_{\rm eff}=\hat{P}\hat{H}\hat{P}+\hat{P}\hat{H}'\hat{S}\hat{H}'\hat{P}.
\end{equation}
After a series simplifications, we can obtain the effective Hamiltonian as
\begin{equation} \label{HamEff1}
	\begin{aligned}
		\hat{H}_{\rm eff}&=\sum_{j=1}^M[U_0+\delta \cos(2\pi\beta j + \phi)]\hat{b}_{j}^{\dag}\hat{b}_j\\&+J^2\left(\frac{2}{U_0+\delta \cos(2\pi \beta j+\phi)}+\frac{ 1}{U_0+\delta \cos(2\pi \beta (j+1)+\phi)}+\frac{1}{U_0+\delta \cos(2\pi \beta (j-1)+\phi)}\right)\hat{b}_{j}^{\dag}\hat{b}_j\\
		&+\sum_{j=1}^MJ^2\left(\frac{ 1}{U_0+\delta \cos(2\pi \beta j+\phi)}+\frac{ 1}{U_0+\delta \cos(2\pi \beta (j+1)+\phi)}\right)(\hat{b}_{j+1}^{\dag}\hat{b}_j+h.c.).
	\end{aligned}
\end{equation}
where $\hat{b}_{j}^{\dag}=\frac{1}{\sqrt{2}}\hat{a}_{j}^{\dag}\hat{a}_{j}^{\dag}$  simultaneously creates two bosons at the $j$th site.
For $|U_0|>>(|J|,|\delta|)$, small variables can be neglected, the effective Hamiltonian becomes   
\begin{equation} \label{HamEff2}
		\hat{H}_{\rm eff}=\sum_{j=1}^M[U_0+\delta \cos(2\pi\beta j + \omega t)]\hat{b}_{j}^{\dag}\hat{b}_j
		+\frac{2 J^2}{U_0}\sum_{j=1}^M(\hat{b}_{j+1}^{\dag}\hat{b}_j+h.c.).
\end{equation}
It turns out to be an AAH model which may support nontrivial topological invariants.

\section{Topological pumping in two particle system}
\begin{figure}
    \centering
    \includegraphics[width=0.95\linewidth]{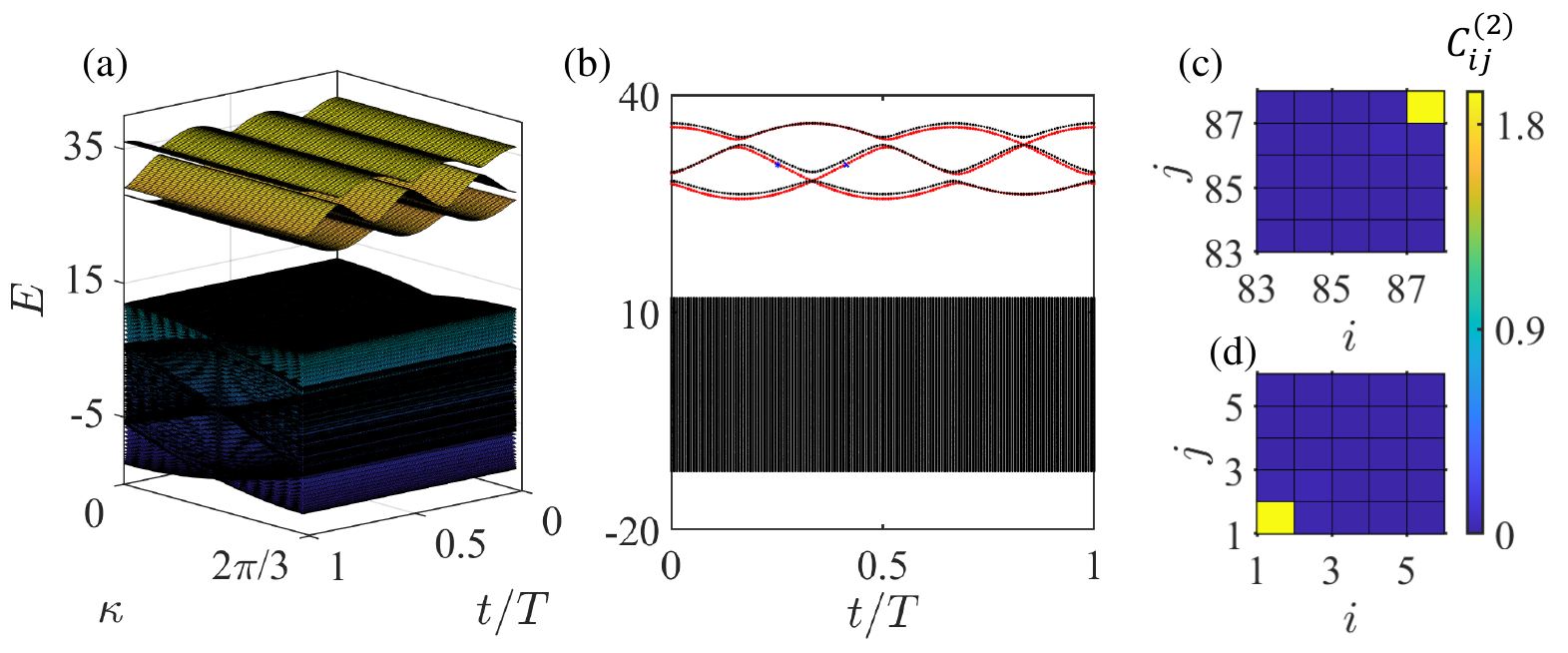}
    \caption{Energy spectrum and edge modes. (a) Multiparticle Bloch bands in 3D view. (b) Energy spectrum as a function of time under the open boundary condition, and the red dots mark the edge modes. (c) and (d): Second-order correlation of two edge states with energies indicated by the blue asterisk and cross in (b), respectively. 
    Parameters are chosen as $N=2, M=87, U_0=30, \delta=5, J=3, \beta=1/3$.}
    \label{fig:2}
\end{figure}
Motivated by the effective Hamiltonians~\eqref{HamEff2}, we may realize interaction-induced topological pumping in two-particle system.
To this end, we first briefly review the theory of multiparticle Thouless pumping~\cite{PhysRevA.95.063630}.
If an interacting system has cotranslation symmetry, that is, the system is invariant if all the particles are shifted by unit cell, we can separate the center-of-mass motion and the relative motion, and find that the center-of-mass (c.m.) momentum $\kappa$ is a good quantum number.
The energies as functions of c.m. momentum form multiparticle Bloch bands, with the corresponding multiparticle Bloch states $|\psi_{m}(\kappa,t)\rangle$.
If the system is periodically modulated by breaking the time-reversal symmetry, the Bloch bands may have nontrivial Chern numbers defined by $|u_{m}(\kappa,t)\rangle$ in the momentum-time space,
\begin{equation}
C_m=\frac{1}{2\pi}\int_0^{\frac{2\pi}{q}}d\kappa\int_0^Tdti(\langle\partial_tu_{m}(\kappa,t)|\partial_{\kappa}u_{m}(\kappa,t)\rangle-\langle\partial_{\kappa}u_{m}(\kappa,t)|\partial_tu_{m}(\kappa,t)\rangle),
\end{equation}
where $T$ is the modulation period and $|u_{m}(\kappa,t)\rangle$ is the cell-periodic part of the Bloch state $|\psi_{m}(\kappa,t)\rangle$.
By preparing a multiparticle Wannier state centered in the $R$th cell,
\begin{equation}
		|W_m(R,t=0)\rangle=\frac{1}{\sqrt{\beta M}}\sum_{\kappa}e^{i\kappa qR}|\psi_{m}(\kappa,0)\rangle,
\end{equation}
the state will adiabatically sweep the occupied multiparticle Bloch band.
In one pumping cycle, the c.m. shift $\Delta P$ of particles is related to the Chern number of the occupied band, $\Delta P= qC_m$. 
In the following, we will first study the topological properties of multiparticle Bloch band, and then show different kinds of topological pumping. 

\subsection{Multiparticle Bloch band and topological invariant}
In individual subspaces of c.m. momentum $\kappa$ under periodic boundary condition, we can obtain eigenvalues by diagonalizing  $\hat{H}(\kappa)|u_{m}(\kappa)\rangle=E_{m}(\kappa)|u_{m}(\kappa)\rangle$, where $\hat{H}(\kappa)$ can be constructed in the basis with translation symmetry and momentum $\kappa$.
The eigenvalues changing with c.m. momentum form Bloch bands.
In Fig.~\ref{fig:2}(a), we show multiparticle Bloch bands as time varies in one period. The parameters are chosen as $N=2, M=87, U_0=30, \delta=5, J=3, \beta=1/3$. 
The energy bands can be classified as two types, that is, the lower continuum bands correspond to scattering states, and the higher three separated bands correspond to bound states.  
The two particles at the same sites form a bound state, due to strong repulsive interaction $U_0>>(|J|,|\delta|)$~\cite{winkler2006repulsively,valiente2008two,PhysRevA.90.062301,fukuhara2013microscopic}. 
Because the scattering-state bands are highly degenerate, the definition of their Chern numbers is vague.
However, different bound-state bands are isolated, and we can calculate the Chern numbers of the three bound-state bands, which turn out to be $C=\{-1,2,-1\}$ from top to bottom. 

The topological properties of the bound-state band can also be related to topological edge states~\cite{PhysRevA.95.053866}.
According to bulk-edge correspondence, if the bulk bands are topologically nontrivial, there should exist topological edge states in the band gaps under open boundary condition.
Fig.~\ref{fig:2}(b) shows the multiparticle energy spectrum under open boundary conditions in the $t-E$ view with the same parameters, and there are edge modes in band gaps marked with red dots.
These edge states are dubbed as topological bound edge states, because the two particles are correlated and localized at left or right ends; see the second-order correlation functions $C_{ij}^{(2)}$ in Fig.~\ref{fig:2}(c) and (d) respectively, where
\begin{equation}
        C_{ij}^{(2)}=\langle \psi | \hat{a}_{i}^{\dag}\hat{a}_{j}^{\dag}\hat{a}_j\hat{a}_i | \psi \rangle.
\end{equation}

\subsection{Thouless pumping of bound state}
\begin{figure}
	\centering
	\includegraphics[width=1\linewidth]{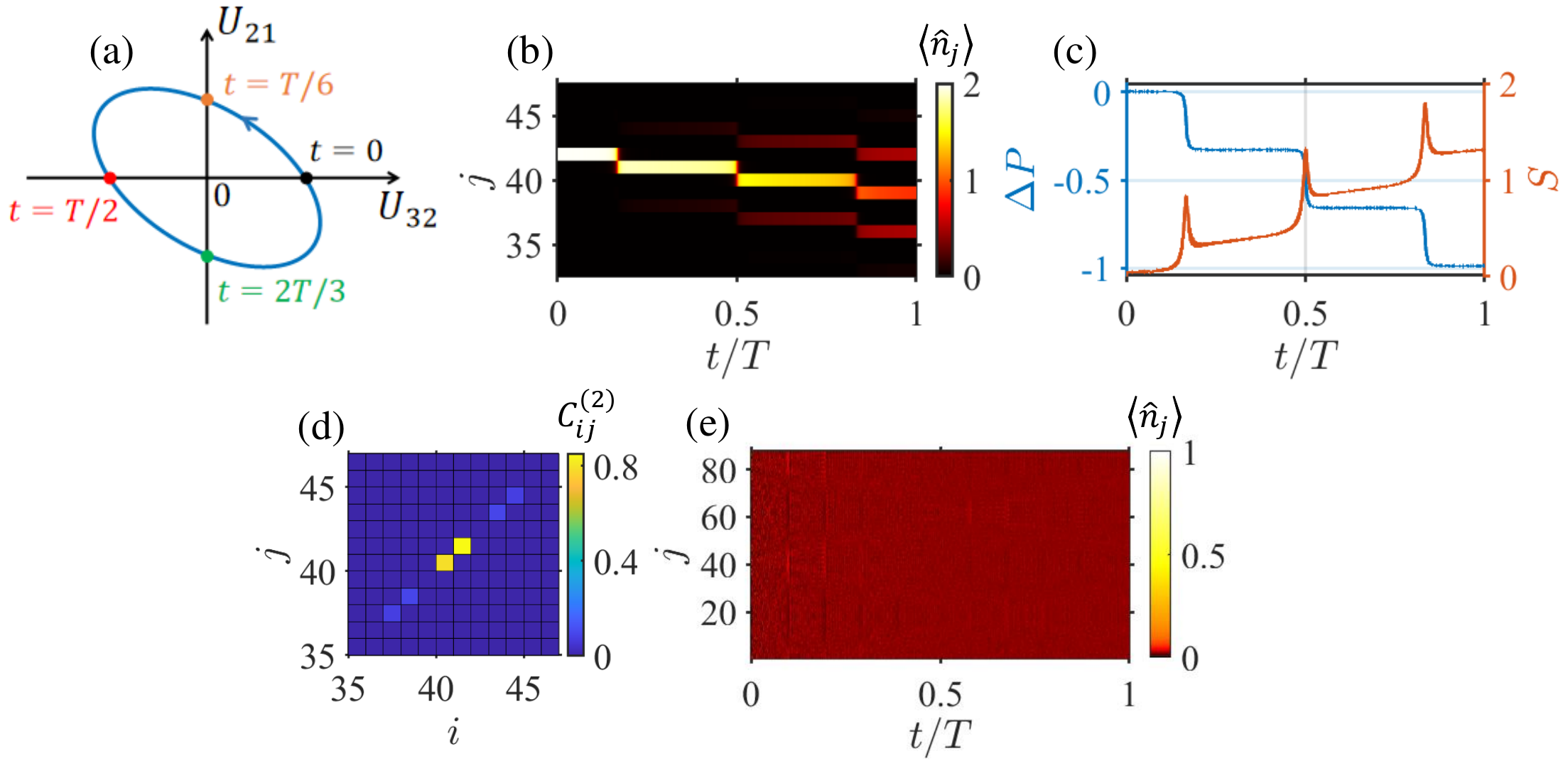}
	\caption{Topological pumping of bound bulk state. (a) The trajectory in the parameter space ($U_{21}\equiv U_{3i+2}-U_{3i+1},U_{32}\equiv U_{3i+3}-U_{3i+2}$) in one pumping cycle. (b) Density distribution as a function of time in a pumping cycle with a bound bulk state $|2\rangle_{42}$ as the initial state. (c) C.m. shift (blue) and instantaneous entanglement entropy (tangerine) between particles as functions of time in one pumping cycle. (d) Second-order correlation of instantaneous state at $t=T/2$. (e) Density distribution as a function of time in a pumping cycle with initial state $|1\rangle_{42}|1\rangle_{82}$, where the two particles are separated. The parameters are set as $U_0=30$, $J=1$, $\delta=2$, $\omega=0.001$ and $M=87$.}\label{fig:3}
\end{figure}

We numerically simulate the Thouless pumping of the two-particle bound state.
Without loss of generality, we consider the bound state $|2\rangle_{42}$ as an initial state, which is almost the maximally localized multiparticle Wannier states~\cite{PhysRevB.56.12847,RevModPhys.84.1419} of the highest multiparticle Bloch bands with Chern number $-1$.  
Note that the middle and lowest bound-state bands with Chern numbers $2$ and $-1$ can also work; see~\ref{Appendix:A}.
The parameters are set as $U_0=30$, $J=1$, $\delta=2$, $\omega=0.001$, and the system size is $87$ sites.  
Dynamics is governed by $|\psi(t)\rangle=\mathcal T \exp[-i\int \hat H(t)dt]|\psi(0)\rangle$, where $\mathcal T$ is a time ordering operator.
Fig.~\ref{fig:3}(a) shows the trajectory in a parameter space per pumping cycle, where $U_{21}=U_{3i+2}-U_{3i+1}$, $U_{32}=U_{3i+3}-U_{3i+2}$, and $i$ can be an integer from $0$ to $M-1$.
Similarly to the previous pumping scheme governed by the Rice-Mele model~\cite{nakajima2016topological,lohse2016thouless}, the trajectory also forms a closed loop that encircles the singular point $(0,0)$, indicating the nontrivial topology of this pumping scheme.
Fig.~\ref{fig:3}(b) shows the time evolution of density distribution $\langle \hat{n}_j(t) \rangle=\langle \psi(t)|\hat{n}_j|\psi(t)\rangle$ for the bound state.
Despite diffusion, the bound state is unidirectionally shifted by $0.9897$ unit cells during a pumping cycle from $t=0$ to $T$, consistent with the corresponding Chern number.
The blue solid line in Fig.~\ref{fig:3}(c) shows the c.m. position shift $\Delta P$ in one cycle, which is defined as 
\begin{equation}
    \Delta P=\frac{1}{q}(\langle \psi(t) | \hat{x} | \psi(t) \rangle - \langle \psi(0) | \hat{x} | \psi(0) \rangle),
\end{equation}
where $\hat{x}=\frac{1}{N}\sum_jj\hat{n}_j$.
During the pumping process, the two bosons are shifted as a whole because of the strong interaction between them.
Fig.~\ref{fig:3}(d) shows the second-order correlation function of state at time $t=T/2$.
We can find that the diagonal line of correlation function is dominated, indicating that the two particles are bound states.
To further illustrate the role of interaction in Thouless pumping, we show the time evolution of density distribution for the initial state $|1\rangle_{42}|1\rangle_{82}$ with the same parameters; see Fig.~\ref{fig:3}(e). 
Obviously, there is no charge pumping in the pumping process and the particles spread over the whole system.
This happens because the interaction and its modulation do not work on the particles that are initially separated.

Another interesting phenomenon is the time evolution of the entanglement entropy~\cite{poshakinskiy2021quantum} between particles, which is defined as  \begin{equation}
  	S=-\sum_{\nu}\lambda_{\nu}ln\lambda_{\nu}.
  \end{equation}
$\lambda_{\nu}$ are the coefficients when performing singular value decomposition (SVD) on a two-particle state $|\psi\rangle=\psi_{i,j}|\psi_{i,j}\rangle$, where $i,j$ represent one of the two particles at the $i$th site and the other at the $j$th site. 
Obviously, due to the exchange symmetry of bosonic particles, swapping positions between any two particles does not change the amplitude, that is, $\psi_{i,j}=\psi_{j,i}$.
$\lambda_{\nu}$ can be obtained by
\begin{equation}
	\psi_{ij}=\sum_{\nu}\sqrt{\lambda_{\nu}}\psi_i^{\nu}\psi_j^{\nu}, 
\end{equation}
with $\sum_{\nu} \lambda_{\nu}=1$. 
The tangerine solid line in Fig.~\ref{fig:3}(c) shows the entanglement entropy of instantaneous states during the pumping process.
We find that the overall trend of the entanglement entropy is increasing along with the diffusion of instantaneous state, and there are peaks near the transition points.
This is because the evolved state at the transition time is bound pair in superposition of two adjacent lattice sites, which can be viewed as a joint effect of both particle and spatial entanglements.

\subsection{Geometrical pumping of bound bloch state}

Geometrical pumping is known as nonquantized transport of a single-momentum state, which is determined by the local geometrical properties of the band structure.
This is in stark contrast to quantization transport in Thouless pumping which is determined by the global topological properties of the filled band~\cite{PhysRevLett.116.200402}.
The pumped charge $Q$ can be calculated using a local current operator $\hat{J}_j(t)$~\cite{PhysRevB.87.085131}, where the density of particles flown through a site $j$ is measured by 
\begin{equation}
    Q^{(j)}(\kappa)=\int_0^T\langle\psi_m(\kappa,t)|\hat{J}_j(t)|\psi_m(\kappa,t)\rangle dt
\end{equation}
in a pumping cycle
$\hat{J}_j(t)$ is obtained via the continuity equation $\hat{J}_j(t)-\hat{J}_{j-1}(t)=i[\hat{n}_j,\hat{H}_M(t)]$, with the form
\begin{equation}
    \hat{J}_j(t)=iJ(\hat{a}^{\dag}_{j}\hat{a}_{j+1}-h.c.).
\end{equation}
Here, we apply the period boundary condition.
Because a unit cell contains three sites, the pumped charge is averaged over one unit by 
\begin{equation}
    Q(\kappa)=\frac{1}{q}\left[Q^{(j)}(\kappa)+Q^{(j+1)}(\kappa)+Q^{(j+2)}(\kappa)\right].
\end{equation}
Here, $Q(\kappa)$ is independent of the lattice site due to the cotranslation symmetry.
$Q(\kappa)$ is generally nonquantized, but the sum of all $Q(\kappa)$ in the Brillouim zone of a certain band gives the quantized pumped charge, 
\begin{eqnarray}
    \frac{1}{q}\sum_j\int_0^T\langle W_m(R,t)|\hat{J}_j|W_m(R,t)\rangle dt
    &=&\frac{1}{q\beta M}\sum_{j,\kappa}\int_0^T\langle \psi_m(\kappa,t)|\hat{J}_j|\psi_m(\kappa,t)\rangle dt\nonumber \\
    &=&\sum_{\kappa}Q(\kappa)=NC_m. \label{Qpumped}
\end{eqnarray}
\begin{figure}
    \centering
    \includegraphics[width=0.9\linewidth]{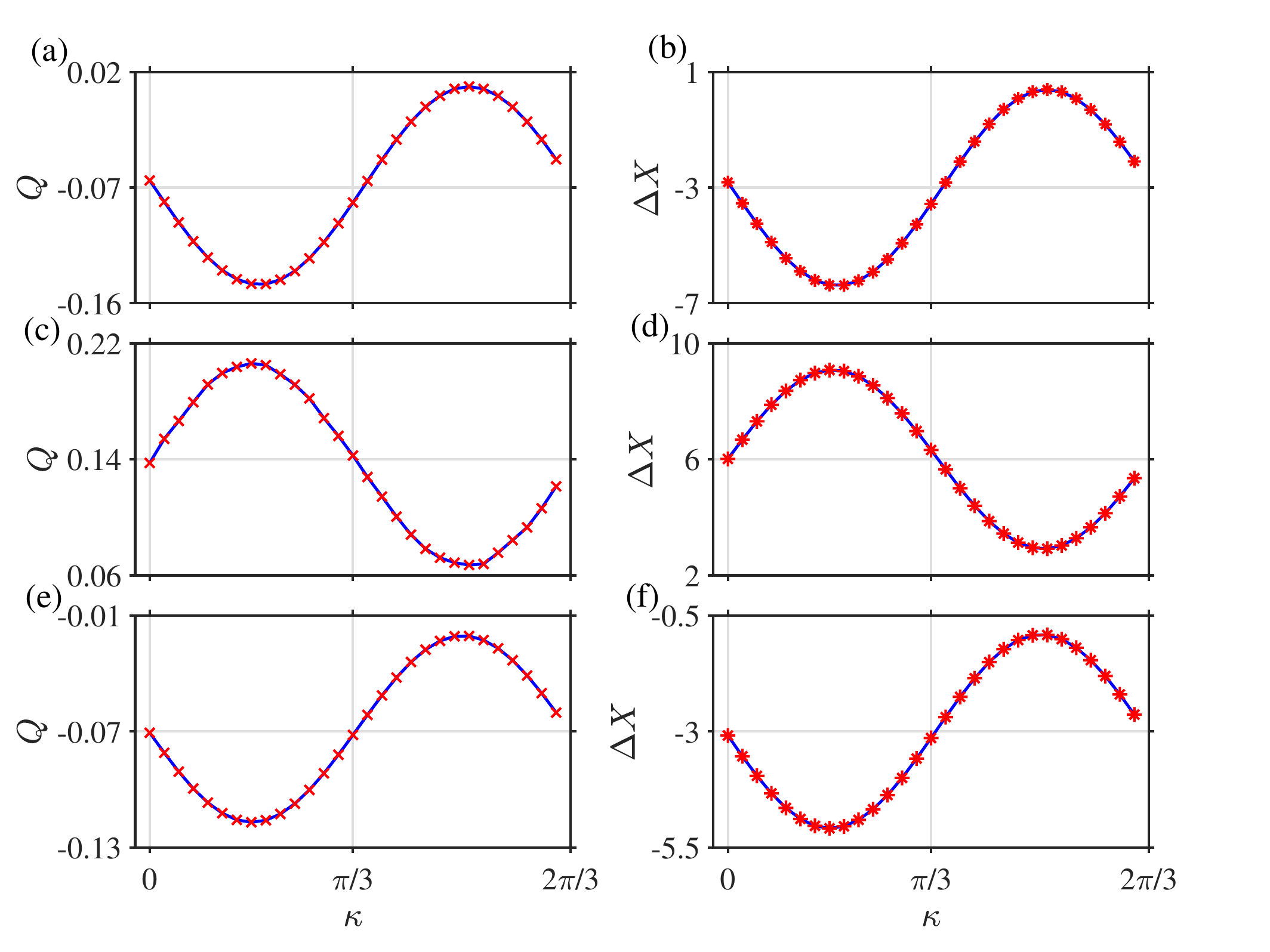}
    \caption{Geometrical pumping of single bound Bloch state. Left panel: Average pumped charge per cycle as a function of c.m. momentum (red cross) for the highest (a), middle (c) and lowest (e) bound-state bands. 
    Right panel: Displacement per cycle as a function of c.m. momentum (red asterisk) for the highest (b), middle (d) and lowest (f) bound-state bands.  The parameters are set as $U_0=30, \delta=2, J=1, \omega=0.001,M=87$ for (a) and (b) and $U_0=30, \delta=5, J=1, \omega=0.0005,M=87$ for (c)-(f).}
    \label{fig:5}
\end{figure}
When the particle number $N$ is equal to $1$, our result returns to the quantized pumping of non-interacting charge~\cite{PhysRevB.27.6083}.
The quantized pumped charge is also consistent with the quantized displacement per cycle.
Similarly to the group velocity of a single-particle momentum state~\cite{RevModPhys.82.1959,PhysRevResearch.5.013020}, the group velocity of a single multiparticle Bloch state is given by
\begin{equation}
    v_m(\kappa,t)=\frac{\partial E_m(\kappa,t)}{\hbar \partial k}+i(\langle\partial_tu_{m}(\kappa,t)|\partial_{\kappa}u_{m}(\kappa,t)\rangle-\langle\partial_{\kappa}u_{m}(\kappa,t)|\partial_tu_{m}(\kappa,t)\rangle).
\end{equation}
Then the displacement of a multiparticle Bloch state per cycle is given by
\begin{equation}
    \Delta X(\kappa)=\int_0^T v_m(\kappa,t) dt.
\end{equation}
Average displacement over the Brillouim zone of the occupied band gives the c.m. shift of the corresponding multiparticle Wannier state 
\begin{equation}
    \frac{q}{2\pi}\int_0^{2\pi/q}\Delta X(\kappa)d\kappa=qC_m. \label{DeltaX}
\end{equation}
Here, the first term of $ v_m(\kappa,t)$ vanishes under the integral of $\kappa$ because of the periodicity of the energy band structure.
Eq.~\eqref{DeltaX} is exactly the same as the formula for multiparticle Thouless pumping~\cite{PhysRevA.95.063630}.
Compared Eq.~\eqref{DeltaX} with Eq.~\eqref{Qpumped}, we can find that both the pumped charge and the displacement per cycle are given by the Chern number of the filled band.

As shown in Fig.~\ref{fig:5}(a), under periodic boundary condition, we consider the initial state as the mutiparticle Bloch state with different $\kappa$ in the highest bound-state band, the pumped charge $Q$ changes with $\kappa$ in the Brillouin zone.
The sum of pumped charge in the Brillouin zone gives $\sum_{\kappa}Q(\kappa)=-1.9898$, consistent with Eq.~\eqref{Qpumped}.
The parameters are set as $U_0=30, \delta=2, J=1, \omega=0.001,M=87$.  
Fig.~\ref{fig:5}(b) shows $\Delta X$ as a function of $\kappa$ with the same condition as Fig.~\ref{fig:5}(a). The average of $\Delta X$ over the Brillouin zone gives $(1/L)\sum_{\kappa}\Delta X(\kappa)=-3$,
which is consistent with Thouless pumping of the initial state as a maximally localized Wannier state uniformly filling the highest bound-state band, where the bound pair is shifted by one unit cell per cycle.
For completeness, Figs.~\ref{fig:5} (c) and (d) respectively show the average pumped charge and displacement per cycle as a function of $\kappa$ with initial states as multiparticle Bloch states of the middle bound band at $t=0$, and the lowest bound band for (e) and (f).
The parameters are $U_0=30, \delta=5, J=1, \omega=0.0005, M=87$, and there is an initial phase $\phi_0=\pi/5$, which is the same as the topological pumping of the bulk states in~\ref{Appendix:A}.
In a pumping cycle, we obtain $
\sum_{\kappa}Q(\kappa)=3.9483$ and $(1/L)\sum_{\kappa}\Delta X(\kappa)=6$ for (c) and (d) respectively; $\sum_{\kappa}Q(\kappa)=-1.9909$ and $(1/L)\sum_{\kappa}\Delta X(\kappa)=-3$ for (e) and (f), respectively.
We find that the profiles of pumped charge and the displacement in geometrical pumping are almost the same, and only differ by a factor related to the particle number and size of a unit cell.

To save computing resource, the numerical simulation of geometrical pumping is performed in a subspace in which the Hamiltonian is expanded only by the basis $\mathcal V=\{|2\rangle_j \}$ and $\mathcal U=\{|1\rangle_j|1\rangle_{j+1}\}$.
The reason is that considering the strong onsite interaction, the instantaneous states during the adiabatic pumping with a two-particle bound state as initial state dominantly occupy the subspace $\mathcal V$, besides $|2\rangle_j$ and $|2\rangle_{j+1}$ are mainly coupled through $|1\rangle_j|1\rangle_{j+1}$.
The validity of this method is numerically verified; see~\ref{Appendix:B}.

\subsection{Topological pumping of bound edge state}
\begin{figure}
    \centering
    \includegraphics[width=1.03\linewidth]{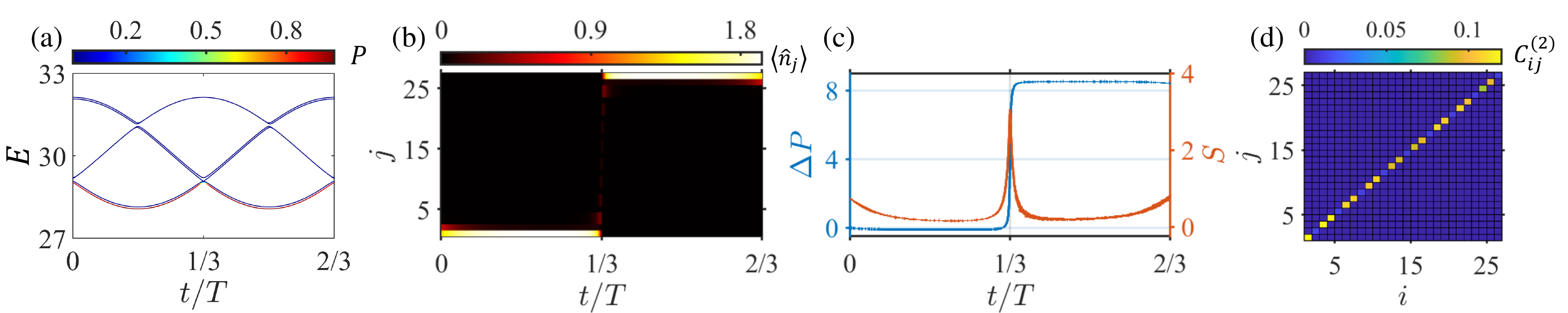}
    \caption{Topological pumping of bound edge state. (a) Energy spectrum of bound states as a function of time under open boundary condition, where colors mark the projection probability of instantaneous state onto eigenstates $P=|\langle \psi_m(t)|\psi(t)\rangle|^2$. The initial state is the edge mode below the lowest bound-state band at $t=0$. (b) Density distribution as a function of time. (c) C.m. shift (blue) and instantaneous entanglement entropy (tangerine) between particles as functions of time. (d) Second-order correlation of instantaneous state at $t=T/3$. The parameters are $U_0=30$, $J=3$, $\delta=2$, $\omega=0.00001$, $M=27$.}
    \label{fig:4}
\end{figure}
We have already shown the topological bound edge states in the band gaps of bound-state bands in previous section.
The energies of the bound edge states connect different bulk bands, and the left bound edge states can be continuously transferred to the right bound edge states through bulk states. 
In this section, we numerically simulate topological pumping of the bound edge state~\cite{PhysRevA.108.032402,olekhno2020topological,tao2023interactioninduced}. 
Without loss of generality, we consider the edge-to-edge transport channel near the lowest bound-state band from $t=0$ to $2T/3$, as shown in Figs.~\ref{fig:4}(a) and (b).
The initial state is the eigenstate below the lowest bound-state band, which can be approximately two bosons at the leftmost site. 
The parameters are set as $U_0=30$, $J=3$, $\delta=2$, $\omega=0.00001$, and $M=27$.   
During the pumping process, the initial left bound edge state is adiabatically transferred to the right side, and the transition occurs at time $t=1/3T$.
Besides,  the entanglement entropy of instantaneous state peaks at the transition time [Fig.~\ref{fig:4}(c)], where the diagonal correlation function has broadest spatial distribution [Fig.~\ref{fig:4}(d)].
The entanglement entropy is nearly the same at the initial and final time because of the same divergence of spatial distribution.

\section{Hybridized pumping in three particle system}
\begin{figure}
	\centering
	\includegraphics[width=0.8\linewidth]{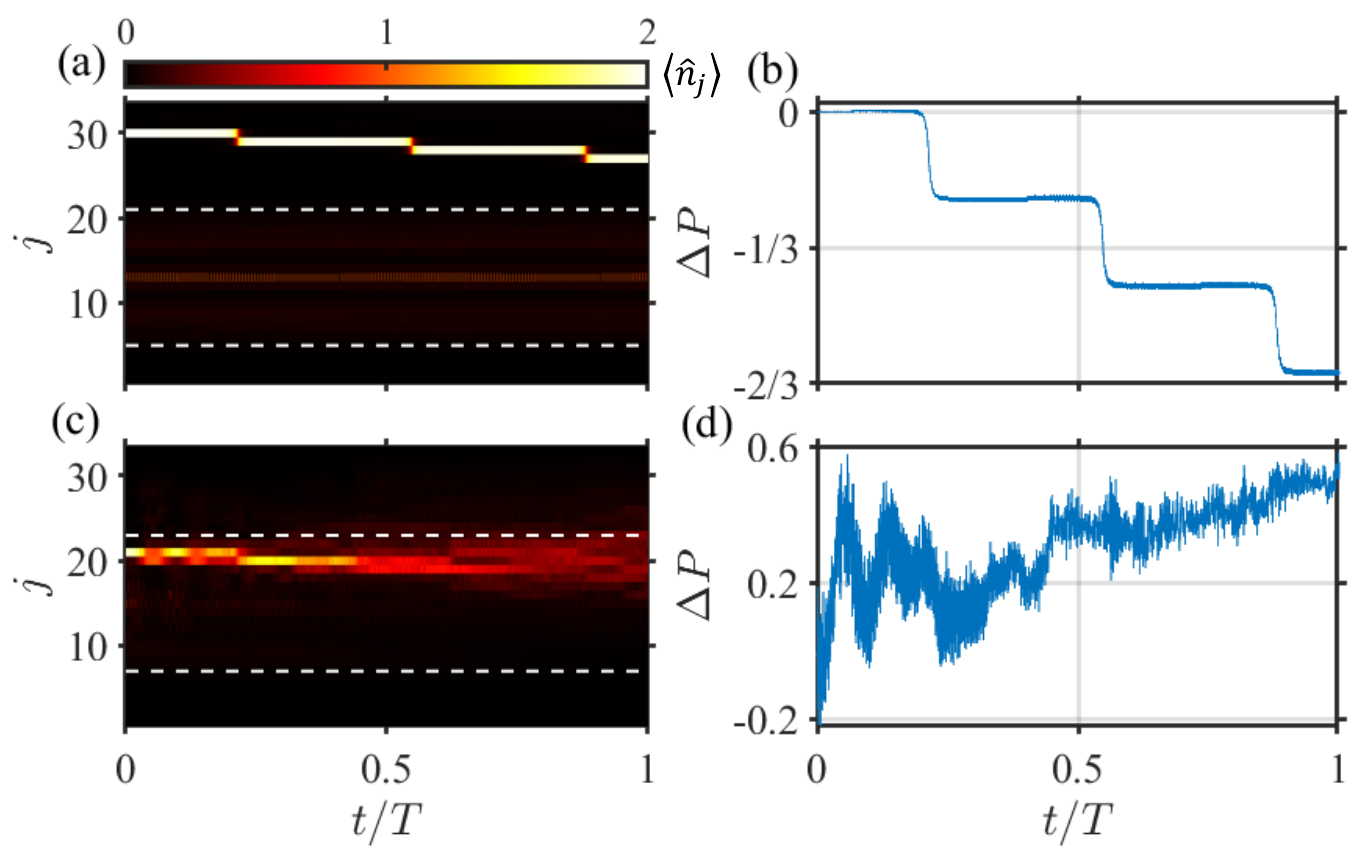}
	\caption{Hybridized fractional pumping and its breakdown. Top: time evolution of (a) density distribution and (b) displacement for the initial state $|1\rangle_{13}|2\rangle_{30}$. Bottom: similar to the top, but the initial state is changed to $|1\rangle_{15}|2\rangle_{21}$.
 The white dashed line denotes the maximal spread width of the single particle with Bloch oscillations. Parameters are set as $U_0=30,J=1,\omega=0.001,\delta=2,F=0.5,M=33$.}
	\label{fig:6}
\end{figure}

In the previous section, we have shown interaction-induced Thouless pumping of two-particle bound states.
Note that a single particle will only undergo quantum walks with zero displacement in our system.
If we consider that a bound pair and a single particle cannot meet each other, we can predict that the overall displacement per particle is $(1\times 2 +0)/3=2/3$ in one cycle, which naturally is a kind of hybridized pumping in few-body systems with fractional c.m. shift per cycle.
Following this similar idea, we can realize hybridized fractional pumping with any designed fractional values by combining more spatially-separated bound pairs and individual particles.
However, the only trouble is that a single particle will spread over the entire lattices without constraint, and the interaction between a single particle and a bound pair is unavoidable.
To overcome this problem, we add an extra term of tilt potential,
\begin{equation}
	\hat{H}_F=\sum_jFj\hat{n}_j.
\end{equation}
Here, $F=\hbar \omega_F$ with $\omega_F$ being the frequency of Bloch oscillations and $\hbar$ is the Planck constant. 
The frequency of Bloch oscillations needs to be a multiple of the driven frequency in Thouless pumping.
Due to the tilted potential, a single particle initially at a single site will undergo Bloch oscillations of breathing modes~\cite{bloch1929quantenmechanik,PhysRevB.46.7252,leo1992observation}.
The single particle will return to its initial position in multiple periods of Thouless pumping.
The maximal spatial spread of the single-particle wavepacket is inversely proportional to the frequency of Bloch oscillations, which is given by~\cite{Hartmann_2004}
\begin{equation}
    L=4J/F.
\end{equation}
In addition, the weak tilt potential does not affect the quantization displacement of the bound pair in the Thouless pump.
We can choose a proper frequency of Bloch oscillations, so that both the adiabatic condition and spatial separation can be satisfied at the same time.

In our simulation, the parameters are set as $U_0=30,J=1,\omega=0.001,\delta=2$, and $F=0.5$, and the size of the lattice system is $M=33$. 
The initial state is $|1\rangle_{13}|2\rangle_{30}$. 
With adiabatic dynamics $|\psi(t)\rangle=\mathcal T \exp[-i\int \hat H(t)dt]|\psi(0)\rangle$, Figs.~\ref{fig:6} (a) and (b) show the time evolution of the density distribution $\langle \hat{n}_j \rangle$ and the c.m. shift $\Delta P$ in a pumping cycle, respectively.
As expected, the response of the two-boson bound state is still a topological pumping with less diffusion, and the behavior of the single one is Bloch oscillations. 
The c.m. position shifts $-0.6454$ unit cells overall, consistent with a $2/3$ fractional shift of displacement, which is attributed to the combination of one unit cell shift of two-boson bound state and that the single particle goes back to its initial position in multiple periods of Bloch oscillations.  
During the pumping process, there is little overlap between the density distribution of the two-boson bound state and the other single particle.

However, if there is spatial overlap between the bound pair and single particle, 
the bound pair and single particle will interact with each other.
The interaction will break both the Thouless pumping of bound pair and the Bloch oscillations of the single particle. 
Figs.~\ref{fig:6}(c) and (d) correspond to the initial state $|1\rangle_{15}|2\rangle_{21}$, where the bound pair and the single particle are quite close.
Because the distance is within the maximal spread width of the single particle,
The overlap between the bound pair and single particle is large.
The parameters are the same as those in Figs.~\ref{fig:6}(a) and (b).
With large density overlap, the phenomenon becomes complex and disordered, and the fractioanl value deviates from $2/3$.
That is why we have to add the tilted potential to separate bound pair and single particle.

\section{Summary and discussion}
We have suggested a form of multiparticle Thouless pumping induced by periodically modulated interaction.
With spatiotenporal modulated onsite interaction in a Bose-Hubbard model, topological pumping of the bound bulk state given by the maximally localized multiparticle Wannier state and bound edge state under open boundary condition can be performed. 
The entanglement entropy between particles also evolves during the pumping process, with a peak value at the transition point related to the dispersion of the single-particle wavepacket.
The topological pumping of bound bulk states can be revealed by a summation of geometrical pumping for different multiparticle Bloch states with different c.m. momenta.
In the next step, it is interesting to explore unexpected richer topological states and phenomena induced by interactions in higher dimensions~\cite{Qin_2018,rachel2018interacting,PhysRevResearch.2.013348,PhysRevA.105.023329,PhysRevA.107.053323}. 
%Increased number of particles is also worth to be considered.
%
Besides, we propose a hybridized pumping with fractional c.m. shift per cycle realized through the combination of Thouless pumping of multiparticle bound state and Bloch oscillations of a single particle.
It is worth studying nontrivial factional topological states in few-body systems~\cite{gemelke2010rotating,PhysRevA.102.063316,PhysRevLett.128.173202,PhysRevResearch.5.013112,leonard2023realization}.

%Except providing a new manner to realize multiparticle Thouless pumping and fractional pumping, our system may motivate more exploration of physical phenomenon under the interaction.

\section{Acknowledgements}{The authors acknowledge useful discussions with Ling Lin, Li Zhang, Na Zhang and Zhoutao Lei. This work was supported by the National Key Research and Development Program of China (Grant No. 2022YFA1404104), the National Natural Science Foundation of China (Grants Nos. 12025509, 12275365), the Key-Area Research and Development Program of Guangdong Province (Grant No. 2019B030330001), and the Natural Science Foundation of Guangdong (Grant No. 2023A1515012099).}

\appendix

\section{Topological pumping of bound states in the middle and lowest bound-state bands}
\label{Appendix:A}
\begin{figure}[!h]
    \centering
    \includegraphics[width=0.8\linewidth]{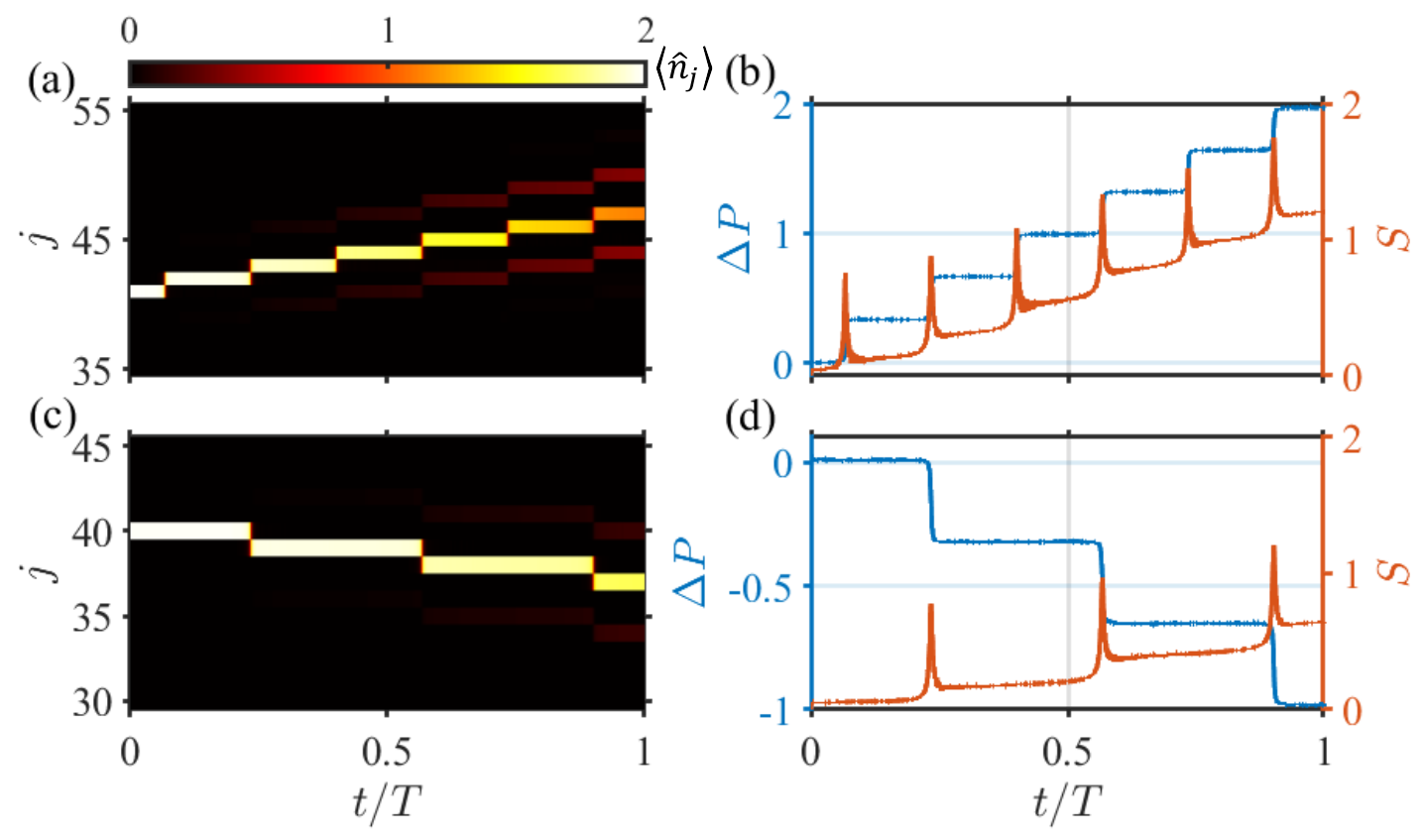}
    \caption{Topological pumping of bound bulk states assisted with the middle and lowest bound-state bands. Density distribution as a function of time in one pumping cycle with an initial phase $\phi_0=\pi/5$. The initial states are set as $|2\rangle_{41}$ for (a) and $|2\rangle_{40}$ for (c); c.m. shift and entanglement entropy as a function of time in the pumping cycle for the initial states $|2\rangle_{41}$ for (b) and $|2\rangle_{40}$ for (d). Parameters are set as $U_0=30, \delta=5, J=1, \omega=0.0005$ and $M=87$.} 
    \label{fig:A1}
\end{figure}
In Figs.~\ref{fig:A1}(a) and (b), we show the time evolution of density distribution, displacement, and entanglement entropy of an initial state $|2\rangle_{41}$.
In one pumping cycle, the c.m. position is shifted by $1.97$ unit cells, and the peaks of entanglement entropy appear at the transition time.
Similar results for the initial state $|2\rangle_{40}$ are shown in Figs.~\ref{fig:A1}(c) and (d), where the c.m. position of particles is shifted by $-0.9728$ unit cells in one cycle. 
The displacements per cycle are consistent with the corresponding Chern numbers, accompanied by an overall increase of the entanglement entropy with peaks emerging at the transition points.

\section{Validity of calculation in subspace} \label{Appendix:B}
\begin{figure}[!h]
    \centering
    \includegraphics[width=0.8\linewidth]{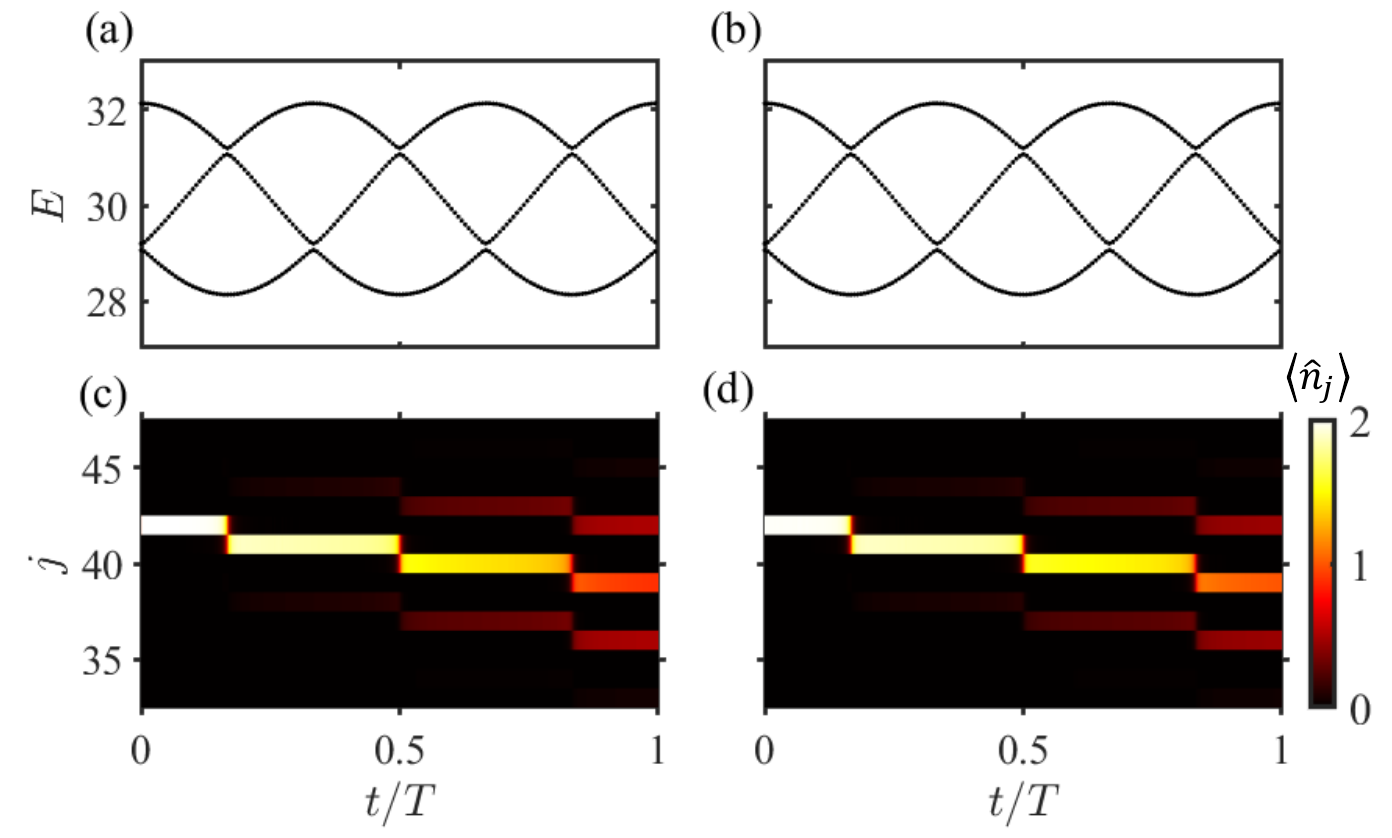}
    \caption{Comparison of energy spectral and density evolution in full Hilbert space and bound state dominated subspace. The bound-state bands shown in $t-E$ view in full Hilbert space for (a) and the subspace for (b); Density distribution as a function of time  with initial state as $|2\rangle_{42}$ simulated in the full space for (c) and subspace for (d). The parameters are chosen as $U_0=30, \delta=2, J=1, \omega=0.001, M=87$.}
    \label{fig:B1}
\end{figure}

We numerically verify the equivalence of simulating the adiabatic evolution of bound states in the total Hilbert space or in a dominated subspace expanded by $\mathcal V=\{|2\rangle_j \}$ and $\mathcal U=\{|1\rangle_j|1\rangle_{j+1} \}$.
Figs.~\ref{fig:B1}(a) and (b) show the multiparticle bound-state band in $t-E$ view for total space and the dominated subspaces, respectively.
Figs.~\ref{fig:B1}(c) and (d) show the density distribution as a function of time in the topological pumping process with initial state $|2\rangle_{42}$, which almost uniformly fills the highest bound-state band. 
The parameters are the same as the corresponding simulation in the main text that $U_0=30, \delta=2, J=1, \omega=0.001, M=87$.
Through the comparison of the two cases, it is obvious that the method of projecting to a reduced dominated subspace is valid and is beneficial for numerical calculation in the many-body region.

\section*{References}
\bibliographystyle{iopart-num}
%\bibliography{reference}
\providecommand{\noopsort}[1]{}\providecommand{\singleletter}[1]{#1}%
\providecommand{\newblock}{}

\end{document}